\documentclass[journal]{IEEEtran}
\ifCLASSINFOpdf
\else
\fi
\usepackage{subfigure}
\usepackage{graphicx}
\usepackage{algorithm}  
\usepackage{algpseudocode}  
\usepackage{amsmath}  
\usepackage{multirow}
\usepackage{cite}
\newcommand{\tabincell}[2]{\begin{tabular}{@{}#1@{}}#2\end{tabular}}
\begin{document}
%
\title{Efficient feature learning and multi-size image steganalysis based on CNN}
%
%
%

\author{Ru~Zhang,
        Feng~Zhu,
        Jianyi~Liu and~Gongshen~Liu,
\thanks{R. Zhang is with School of Cyberspace Security, Beijing University Posts and Telecommunications, Beijing, China(email:zhangru@bupt.edu.cn)}
\thanks{F. Zhu is with School of Cyberspace Security, Beijing University Posts and Telecommunications, Beijing, China(email:zhufeng@bupt.edu.cn)}
\thanks{J. Liu is with School of Cyberspace Security, Beijing University Posts and Telecommunications, Beijing, China(email:liujianyi1987@163.com)}
\thanks{G. Liu is with School of Electronic Information and Electrical Engineering, Shanghai Jiao Tong University, Shanghai, China(email:lgshen@sjtu.edu.cn)}

}
\maketitle


\begin{abstract}
  For steganalysis, many studies showed that convolutional neural network has better performances than the two-part structure of traditional machine learning methods. However, there are still two problems to be resolved: cutting down signal to noise ratio of the steganalysis feature map and steganalyzing images of arbitrary size. 
  Some algorithms required fixed size images as the input and had low accuracy due to the underutilization of the noise residuals obtained by various types of filters. In this paper, we focus on designing an improved network structure based on CNN to resolve the above problems.
  First, we use $3\times 3$ kernels instead of the traditional $5\times 5$ kernels and optimize convolution kernels in the preprocessing layer. The smaller convolution kernels are used to reduce the number of parameters and model the features in a small local region. Next, we use separable convolutions to utilize channel correlation of the residuals, compress the image content and increase the signal-to-noise ratio (between the stego signal and the image signal).
  Then, we use spatial pyramid pooling (SPP) to aggregate the local features, enhance the representation ability of features, and steganalyze arbitrary size image. Finally, data augmentation is adopted to further improve network performance.
  The experimental results show that the proposed CNN structure is significantly better than other four methods such as SRM, Ye-Net, Xu-Net, and Yedroudj-Net, when it is used to detect two spatial algorithms such as WOW and S-UNIWARAD with a wide variety of datasets and payloads.
\end{abstract}


%
\IEEEpeerreviewmaketitle

\section{Introduction}
%
%
%
%
\IEEEPARstart{F}{or} a long time, steganography and steganalysis always developed in the struggle with each other. Steganography seeks to hide secret information into a specific cover as much as possible and makes the changes of cover as little as possible, so that the stego is close to the cover in terms of visual quality and statistical characteristics[1,2,3]. Meanwhile, steganalysis uses signal processing and machine learning theory, to analyze the statistical differences between stego and cover. It improves detecting accuracy by increasing the number of features and enhancing the classifier performance[4].

\par Currently, the existing steganalysis methods include specific steganalysis algorithms and universal steganalysis algorithms. Early steganalysis methods aimed at the detection of specific steganography algorithms[5], and the general-purpose steganalysis algorithms usually use statistical features and machine learning[6]. The commonly used statistical features include the binary similarity measure feature[7], DCT[8,9] and wavelet coefficient feature[10], co-occurrence matrix feature[11] and so on. In recent years, higher-order statistical features based on the correlation between neighboring pixels have become the mainstream in the steganalysis. These features improve the detection performance by capturing complex statistical characteristics associated with image steganography, such as SPAM[12], Rich Models[13], and its several variants[14,15]. However, those advanced methods are based on rich models that include tens of thousands of features. Dealing with such high-dimensional features will inevitably lead to increasing the training time, overfitting and other issues. Besides, the success of feature-based steganalyzer to detect the subtle changes of stego largely depends on the feature construction. The feature construction requires a great deal of human intervention and expertise.
\par Benefiting from the development of deep learning, convolutional neural networks (CNN) perform well in various steganalysis detectors[16,17,18,19]. CNN can automatically extract complex statistical dependencies from images and improve the detection accuracy. Considering the GPU memory limitation, existing steganography analyzers are typically trained on relatively small images (usually $256\times 256$). But the real-world images are of arbitrary size. This leads to a problem that how an arbitrary sized image can be steganalyzed by the CNN-based detector with a fixed size input. In traditional computer vision tasks, the size of the input image is usually adjusted directly to the required size. However, this would not be a good practice for steganalysis as the relation between pixels are very weak and independent. Resizing before classification would compromise the detector accuracy.
\par In this paper, we have proposed a new CNN network structure named ``Zhu-Net'' to improve the accuracy of spatial domain steganalysis. The proposed CNN performs well in both the detection accuracy and compatibility, and shows some distinctive characteristics compared with other CNNs, which are summarized as follows:
\par (1) In the preprocessing layer, we modify the size of the convolution kernel and use 30 basic filters of SRM[13] to initialize the kernels in the preprocessing layer to reduce the number of parameters and optimize local features. Then, the convolution kernel is optimized by training to achieve better accuracy and to accelerate the convergence of the network.
\par (2) We use two separable convolution blocks to replace the traditional convolution layer. Separable convolution can be used to extract spatial correlation and channel correlation of residuals, to increase the signal to noise ratio, and obviously improve the accuracy.
\par (3) We use spatial pyramid pooling[20] to deal with arbitrary sized images in the proposed network. Spatial pyramid pooling can map feature maps to fixed lengths and extract features through multi-level pooling. 
\par We design experiments to compare the proposed CNN network with Xu-Net[17], Ye-Net[19], and Yedroudj-Net[21]. The proposed CNN shows excellent detection accuracy, which even exceeds the most advanced manual feature set, such as SRM[13].
\par The rest of the paper is organized as follows. In Section II, we present a brief review of the framework of popular image steganalysis methods based on convolutional neural networks (CNNs) in the spatial domain. The proposed CNN is described in Section III, which is followed by experimental results and analysis in Section IV. Finally, the concluding remarks are drawn in Section V.

\section{Related Works}

\par The usual ways to improve CNN structure for steganalysis include: using truncated linear units, modifying topology by mimicking the Rich Models extraction process, and using deeper networks such as ResNet[22], DenseNet[23], and others.
\par Tan et.al used a CNN network with four convolution layers for image steganalysis[24]. Their experiments showed that a CNN with random initialized weights usually cannot converge and initializing the first layer's weights with the KV kernel can improve accuracy. Qian et al.[25] proposed a steganalysis model using standard CNN architecture with Gaussian activation function, and further proved that transfer learning is beneficial for a CNN model to detect a steganography algorithm with low payloads. The performance of these schemes is comparable to or better than the SPAM scheme[12], but is still worse than the SRM scheme[13]. Xu et al.[17] proposed a CNN structure with some techniques used for image classification, such as batch normalization (BN)[26], 1×1 convolution, and global average pooling. They also did pre-processing with a high-pass filter and used an absolute (ABS) activation layer. Their experiments showed better performance. By improving the Xu-CNN, they achieved a more stable performance[27]. In JPEG domain, Xu et al.[18] proposed a network based on decompressed image and achieved better detection accuracy than traditional methods in JPEG domain. By simulating the traditional steganalysis scheme of hand-crafted features, Fridrich et al.[28] proposed a CNN structure with histogram layers, which is formed by a set of Gaussian activation functions. Ye et al.[19] proposed a CNN structure with a group of high-pass filters for pre-processing and adopted a set of hybrid activation functions to better capture the embedding signals. With the help of selection channel knowledge and data augmentation, their model obtained significant performance improvements than the classical SRM. Fridrich[29] proposed a different network architecture to deal with steganalyzed images of arbitrary size by manual feature extraction. Their scheme inputs statistical elements of feature maps to the fully-connected-network classifier.
\par Generally, there are two disadvantages for the existing networks.
\par (1) A CNN is composed of two parts: the convolution layer and the fully connected layer (ignoring the pooling layer, etc.). The function of convolution layer is to convolve input and to output the corresponding feature map. The input of the convolution layer does not need a fixed size image, but its output feature maps can be of any size. The fully connected layer requires a fixed-size input. Hence, the fully connected layer leads to the fixed size constraint for network. The two existing solutions are as follows.
\begin{itemize}
  \item Resizing the input image directly to the desired size. However, the relationship between the image pixels is fragile and independent in the steganalysis task. Detecting the presence of steganographic embedding changes really means detecting a very weak noise signal added to the cover image. Therefore, resizing the image size directly before inputting image to CNN will greatly affect the detection performance of the network.
  \item Using a full convolutional neural network(FCN), because the convolutional layer does not require a fixed image size.
  \end{itemize}
\par In this paper, we propose the third solution: mapping the feature map to a fixed size before sending it to the fully-connected layer, such as SPP-Net[20]. The proposed network can map feature maps to a fixed length by using spp-module, so as to steganalyze arbitrary size images.

\par (2) Accuracy of steganalysis based on CNN seriously relies on signal-to-noise ratio of feature maps. CNN network favorites high signal-to-noise ration to detect small differences between stego signals and cover signals. Many steganalyzers usually extract the residuals of images to increase the signal-to-noise ratio. However, some existing schemes directly convolve the extracted residuals without thinking of the cross-channel correlations of residuals, which do not make good use of the residuals.
\par In this paper, we increase signal-to-noise ratio by three ways as follows.
\begin{itemize}
  \item Optimizing the convolution kernels by reducing kernel size and the proposed ``forward-backward-gradient descent'' method.
  \item Using group convolution to process the spatial correlation and channel correlation of residuals separately. 
\end{itemize}
\par We greatly improve the accuracy of steganalysis by combining the above two ways.

\section{Proposed Scheme}

\begin{figure*}
	\includegraphics[scale=0.6]{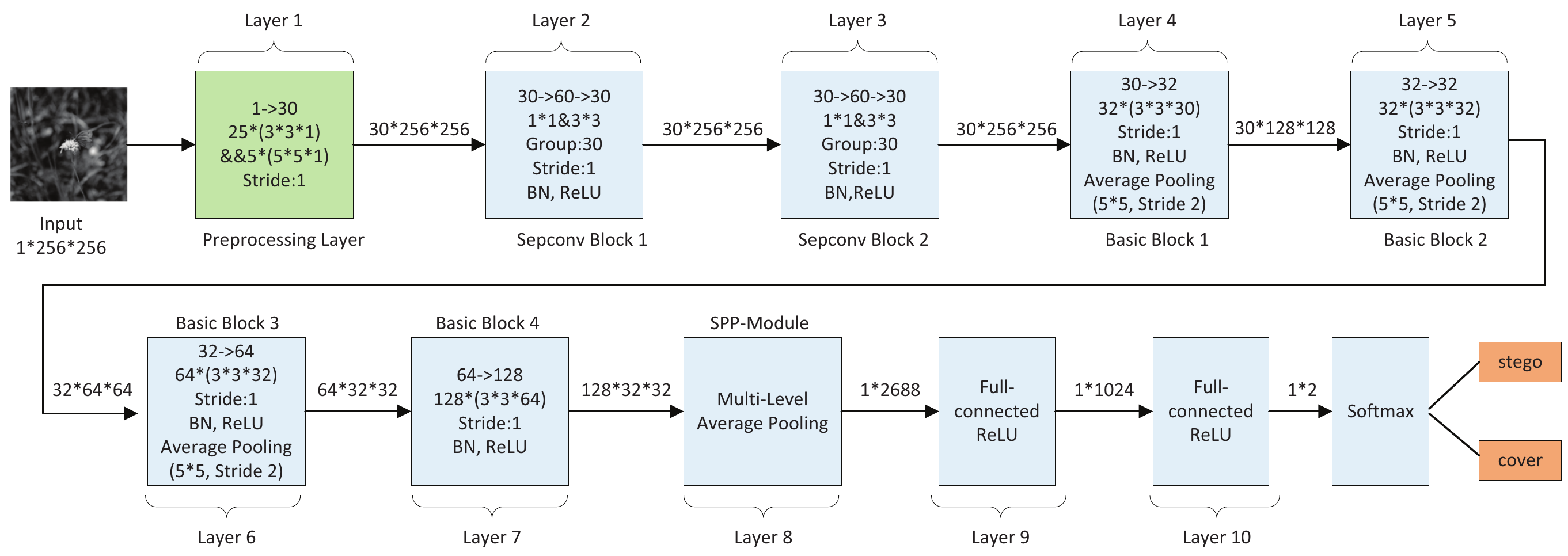}
	\caption{The architecture of the proposed CNN. For each block, $x_1 \rightarrow x_2;x_2 (a*a*x_1)$ denotes the block with the kernel size $a*a$ for $x_1$ input feature maps and $x_2$ output feature maps. Batch normalization is abbreviated as BN.}

\end{figure*}

\subsection{Architecture}
\par The framework of the proposed CNN based steganalysis is shown in Fig. 1. The CNN accepts an input image of size $256 \times 256$ and outputs a two class labels (stego and cover). The proposed CNN is composed of a number of layers including one image preprocessing layer, two separable convolution (sepconv) block, four basic blocks for feature extraction, a spatial pyramid pooling(SPP) module, and two fully connected layers followed by a softmax.
\par The convolutional blocks have four blocks marked as `Basic Block 1' through `Basic Block 4' to extract spatial correlation between feature maps and finally transport to the fully connected layer for classification. Each Basic Block is made of the following steps:

\subsubsection{Convolution Layer}
\par Unlike the existing networks that use large convolution kernel(e.g. $5\times 5$), we use small convolution kernels(e.g. $3\times 3$) to reduce the number of parameters. 
The small convolution kernels can increase the nonlinearity of the network, which significantly increase the capability of feature representation. 
Therefore, we set the size of the convolutional kernels to $3\times 3$ for the Basic Block 1-4. For Basic Block 1 to Basic Block 4, there are 32, 32, 64, 128 channels. The number of channels in each Basic Block is also based on a comprehensive consideration of computational complexity and network performance. Stride and padding size are shown in the Fig. 1.

\subsubsection{Batch Normalization (BN) layer}
\par The Batch normalization(BN)[26] is usually used to normalize the distribution of each mini-batch to a zero-mean and a unit-variance during the training. The advantage of using a BN layer is that it effectively prevents the gradient vanishing/exploding and over-fitting in the deep neural network[26], and allows a relatively large learning rate to speed up the convergence. From the experiments, we found that the networks without BN, such as Ye-Net, are very sensitive to the initialization of parameters and may not converge with inappropriate initializations. Therefore, we use BN in the proposed scheme.

\subsubsection{Non-linear activation function}

\par For all the blocks in Zhu-Net, we use the classical rectifying linear unit (ReLU) as the activation function to prevent gradient vanishing/exploding, produce sparse features, accelerate network convergence and so on. Applying ReLU to neurons can make them selectively respond to useful signals among the inputs, resulting in more efficient features. The ReLU function is also convenient for derivation and benefits back-propagation gradient calculations. We do not use the truncated linear unit (TLU) in our network, because we find that the TLU decreases the non-linearity. To verify that, we compare TLU (threshold $T=3$) with ReLU. From the Table \uppercase\expandafter{\romannumeral1}, Zhu-Net with ReLU has lower error rate of detecting various steganography algorithms. ReLU also accelerates the convergence and shows better performance than TLU, as shown in Fig.2.

\begin{table}
	\centering
	\caption{Steganalysis error rates comparison of Zhu-Net with TLU and Zhu-Net with ReLU against two algorithms WOW and S-UNIWARD at 0.2 bpp and 0.4 bpp. Both networks are trained and tested on BOSS dataset.}
	\begin{tabular}{ccc}

		\hline
		Algorithms& \tabincell{c}{Zhu-Net with \\TLU}& \tabincell{c}{Zhu-Net with \\ReLU}\\
		\hline
		WOW(0.2bpp)       & 0.257 & \textbf{0.233} \\
		WOW(0.4bpp)       & 0.138 & \textbf{0.118} \\
		S-UNIWARD(0.2bpp) & 0.316 & \textbf{0.285} \\
		S-UNIWARD(0.4bpp) & 0.188 & \textbf{0.153} \\
		\hline
	\end{tabular}
\end{table}

\begin{figure}
	\centering
	\includegraphics[width=0.5\textwidth]{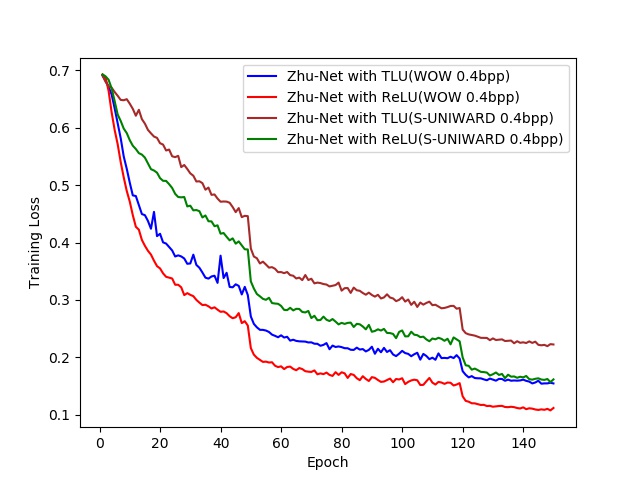}
	\caption{Comparing convergence performances of training Zhu-Net with TLU and Zhu-Net with ReLU against two algorithms WOW and S-UNIWARD at 0.2 bpp and 0.4 bpp. Both networks are trained and tested on BOSS dataset.}
\end{figure}

\subsubsection{Average pooling layer}

\par Average pooling layers are used in Basic Block 1 to Basic Block 3. It down-samples feature maps, better abstracts the image features, reduces the size of feature map and enlarges the receptive fields. With invariance, the average pooling also enhances the generalization ability of the network.

\par Note that we design separable convolution blocks to enhance SNR (signal noise ratio of stego signal to image) and remove the image content effectively from features. In the last block, we use a SPP module to better extract features. The SPP module enriches feature expressions by multi-level pooling, such that our network can train and test with multi-size images. Details are elaborated in Section III-D.  

\subsection{Improving Kernels}

\par The embedding operation of steganography can be viewed as adding a smaller amplitude noise signal to the cover signal. Therefore, it is a good idea to perform residual calculation prior to feature extraction in network. In the preprocessing layer, we use a set of high-pass filter (for example, 30 basic high-pass filters of SRM[12], that is, the “spam” filters and their rotated counterparts, Similar to Ye-Net [19] and Yedroudj-Net[21]) to extract noise residuals map from input image. In [24], the authors showed that without such preliminary high-pass filter, CNN convergence rate should be very slow. Hence, using multiple filters can effectively improve network performance.
\par We use the following strategy to initialize the weights of the preprocessing layer.

\subsubsection{Small sized kernels}
\par Small sized convolution kernels can reduce the number of parameters and prevent modeling larger regions, to effectively reduce calculations. As some existing schemes showed, the kernel size of  $5\times 5$ is suitable for some filters in SRM, such as ``SQUARE  $5\times 5$'', ``EDGE  $5\times 5$''. But for the remaining 25 filters, the  $5\times 5$ convolution kernel will model residual elements in a big local region. So we keep ``SQUARE  $5\times 5$'' and ``EDGE  $5\times 5$'', and use  $3\times 3$ size for the remaining 25 high-pass filters. We initialize the central part of convolution kernels with the SRM kernels and pad the remaining elements to zero, as shown in Fig. 3. Two parts of the residuals calculated by these filters are stacked together as the input of the next convolutional layer.
\begin{figure}
	\centering
	\includegraphics[scale=0.6]{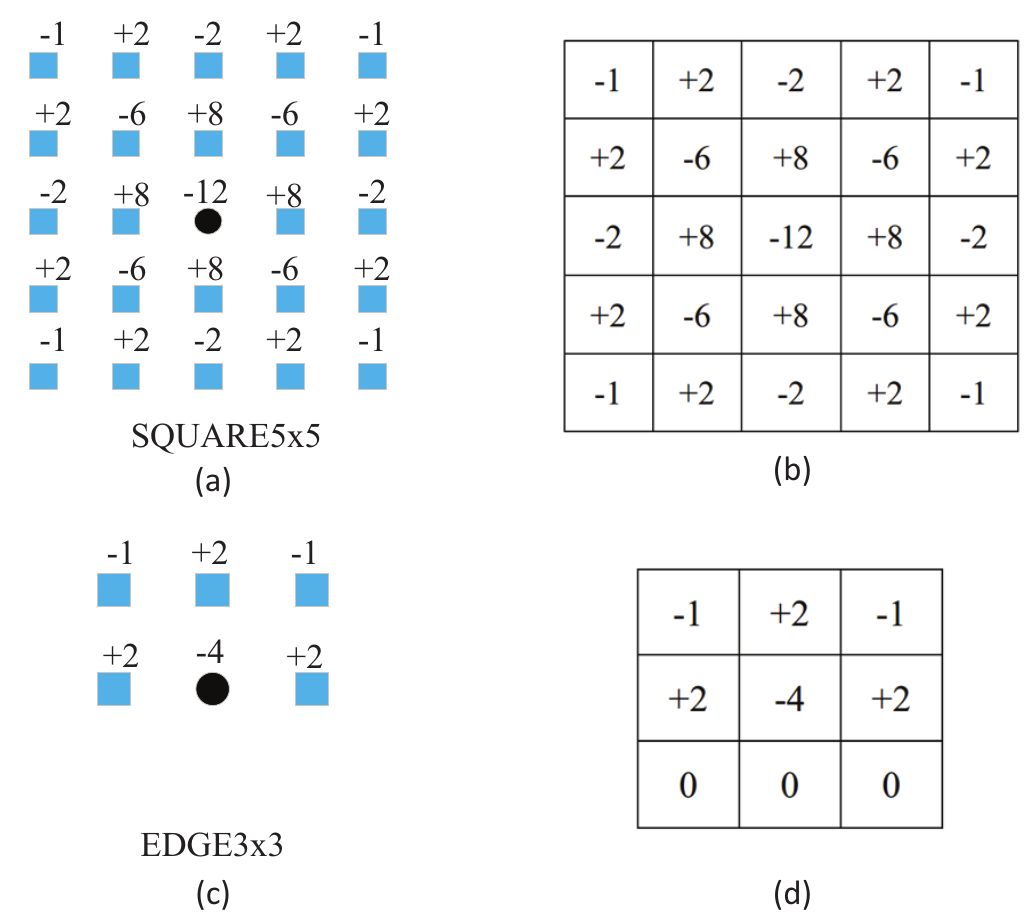}
	\caption{An example of initializing convolution kernels wiht different sized high-pass filters. (a) SQUARE $5\times 5$ high-pass filter. (b) The convolution kernel corresponding to (a). (c) EDGE$3\times 3$ high-pass filter. (d) The convolution kernel corresponding to (c).}

\end{figure}

\subsubsection{Optimizing kernels}
\par Modeling residuals instead of pixel values can extract more robust features. In Yedroudj-Net and Xu-net, convolution kernels in the preprocessing layer are fixed during the training. To optimize the SRM hand-crafted feature set designed with domain knowledge, we design a method named ``forward-backward-gradient descent'', and use it in preprocessing layer. We calculate the residual as follows:
\par For each image $X = X_{ij}$, the residual $R = R_{ij}$ is:
\begin{equation}
	R_{ij} = X_{pred}(N_{ij}) - cX_{ij},
\end{equation}

\par where $c \in N$ is the residual order, $N_{ij}$ is the neighboring pixels of $X_{ij}$ and $X_{pred}(.)$ is a predictor of $cX_{ij}$ defined on $N_{ij}$. In practice, we usually use high-pass filters to achieve $X_{pred}(.)$.

\par The complete process of optimizing the kernel is given as follows:

\emph{\textbf{The forward-backward-gradient descent Method}}

\emph{\textbf{Forward propagation：}}
\par \textbf{Input:} An image $X = X_{ij}$, the high-pass filter $K$
\par \textbf{Output:} The noise residuals map $R = R_{ij}.$
\par \textbf{Step 1:} Initialization: The convolution kernel of the preprocessing layer is initialized by a high-pass filter in the SRM, and the weights of the convolution kernel are described by $K$.
\par \textbf{Step 2:} Calculate residuals:
\begin{equation}
	R=X\ast K = (\sum _{m,n}X_{i,j}^{m,n}\cdot K^{m,n}),
\end{equation}

\par where $\ast$ denotes the convolution operator, and $m, n$ are the corresponding index of the kernel $K$.

\emph{\textbf{Back propagation：}}
\par \textbf{Input:} The gradient of the previous layer $\delta ^{l+1}$, the high-pass filter $K$.
\par \textbf{Output:} The gradient of the preprocessing layer $\delta ^{l}.$
\par \textbf{1:} Let the backward gradient of the previous layer be $\delta ^{l+1}$. Then the gradient of the preprocessing layer is:
\begin{equation}
	\delta ^{l} = \frac{\partial Loss}{\partial K} = \frac{\partial Loss}{\partial R}\frac{\partial R}{\partial K} = \delta ^{l+1} * K,
\end{equation}
\par \textbf{2:} Return the gradient of the preprocessing layer $\delta ^{l+1}$.

\emph{\textbf{Gradient descent:}}
\par \textbf{Input:} The gradient of the preprocessing layer $\delta ^{l}$, the high-pass filter $K$, the learning rate $lr$.
\par \textbf{Output:} The optimized kernels $K^{'}.$
\par \textbf{1:} Optimize the weight of the preprocessing layer by:
\begin{equation}
	K^{'} = K - lr * \delta^{l},
\end{equation}
\par \textbf{2:} Return the optimized kernels $K^{'}$.

\par The corresponding experiment results are shown in Fig. 4 and Table \uppercase\expandafter{\romannumeral2}. We compared the Zhu-Net with fixed kernels, and the network Zhu-Net with optimizable kernels.

\begin{table}
	\centering
	\caption{Steganalysis error rates comparison between Zhu-Net with fixed kernels and Zhu-Net with optimized kernels against two steganography algorithms WOW and S-UNIWARD at 0.2 bpp and 0.4 bpp. Both networks are trained and tested on BOSS dataset. }
	\begin{tabular}{ccc}

		\hline
		Algorithms& \tabincell{c}{Zhu-Net with \\fixed kernels}& \tabincell{c}{Zhu-Net with \\optimized kernels}\\
		\hline
		WOW(0.2bpp)       & 0.243 & \textbf{0.233} \\
		WOW(0.4bpp)       & 0.130 & \textbf{0.118} \\
		S-UNIWARD(0.2bpp) & 0.324 & \textbf{0.285} \\
		S-UNIWARD(0.4bpp) & 0.169 & \textbf{0.153} \\
		\hline
	\end{tabular}
\end{table}

\begin{figure}
	\centering
	\includegraphics[width=0.5\textwidth]{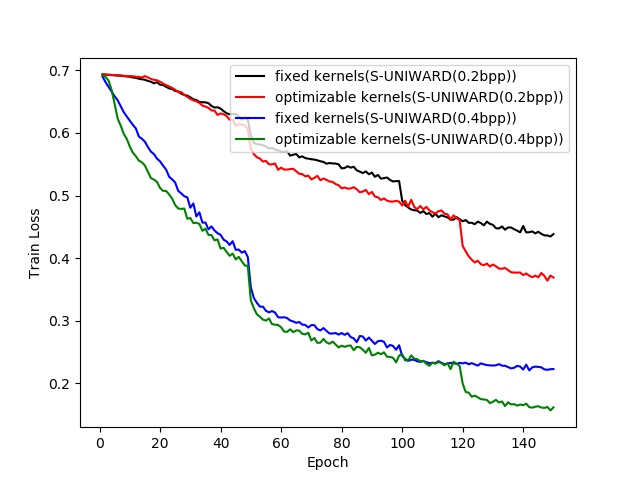}
	\caption{Comparing convergence performances of training Zhu-Net with fixed kernels and Zhu-Net with optimized kernels against two algorithms WOW and S-UNIWARD at 0.2 bpp and 0.4 bpp. Both networks are trained and tested on BOSS dataset.}
\end{figure}
\par From Table II, it is observed that using the “forward-backward-optimize” method, Zhu-Net achieves higher accuracy than the network with fixed kernels for detecting various steganography algorithms. According to Fig. 4, our network is more quickly converged and has less training loss than the network with fixed kernels.


%

\subsection{Separable Convolution}
\subsubsection{Problem Formulation}
\par Existing steganalysis schemes directly learn filters in a 3D space without thinking of cross-channel correlations of the residuals, so that the residual information is not well utilized. To resolve this problem, we used two separable convolution blocks (i.e. the sepconv blocks) consisting of a $1\times 1$ convolution and a $3\times 3$ convolution after preprocessing the layer (as shown in Fig. 1). 
\par Separable convolution has recently made great progress in computer vision tasks, such as Inception[30], Xception[31] and other structures. Xception, a variant of an Inception module is shown in Fig. 5(a). This extreme version of inception completely separates the correlation between channels, reducing storage space and enhancing the expressiveness of the model. Therefore, we use Xception structure to design the corresponding sepconv block to achieve the group convolution of residuals.

\begin{figure}
	\centering
	\includegraphics[width=0.5\textwidth]{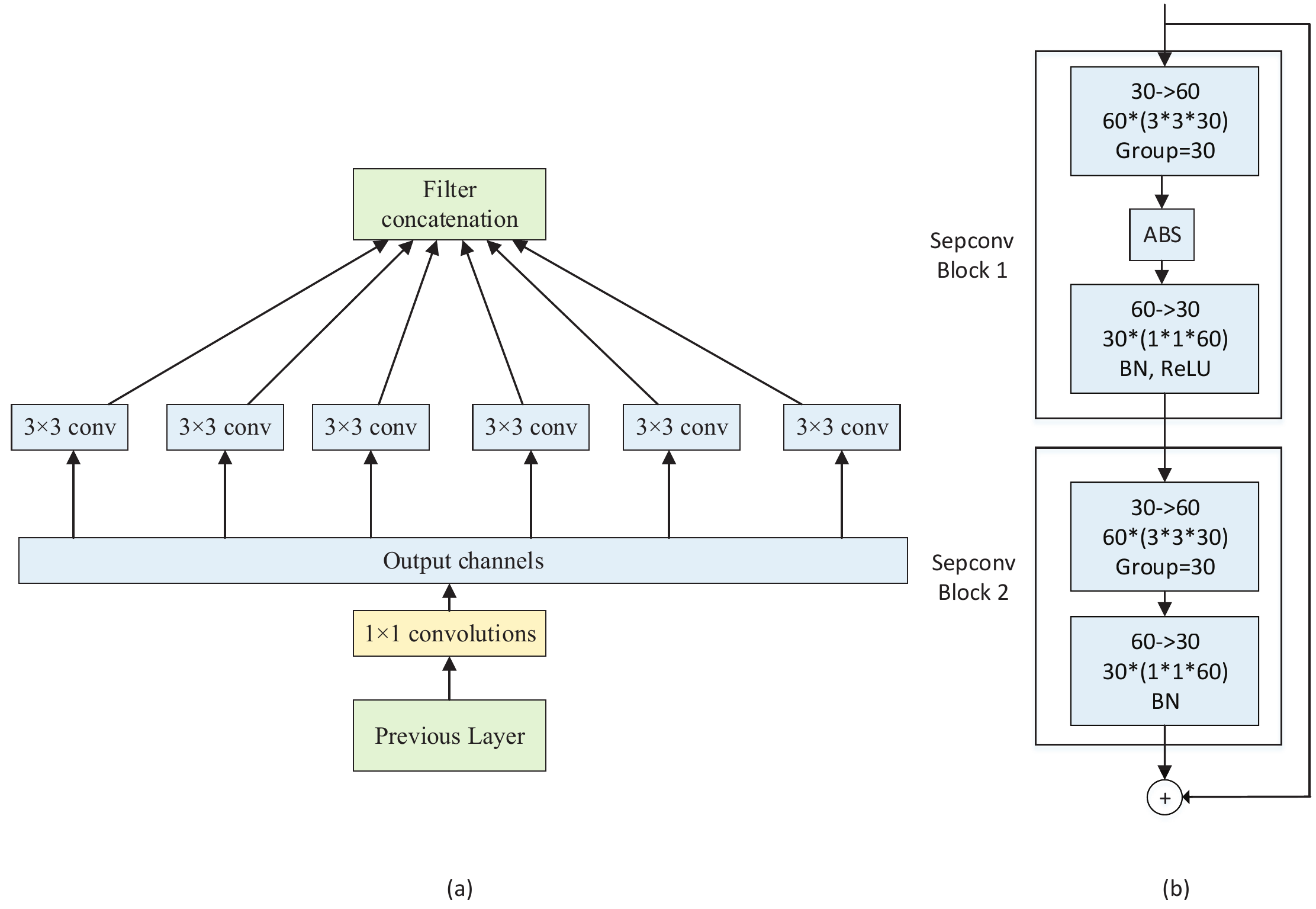}
	\caption{ (a) A variant version of Inception module[31]; (b) structure of sepconv blocks}
\end{figure}

\par In our scheme, we assume that the channel correlation and spatial correlation of residuals are independent. The sepconv block can perform group convolution for each feature map generated by a high-pass filter, which makes full use of the residual information and removes image contents from features to improve the signal-to-noise ratio. The design of the sepconv blocks is shown in Fig. 5(b).
\par First, $1\times 1$ pointwise convolution is performed in a sepconv block to extract residual channel correlations. Then $3\times 3$  depthwise convolution is performed to extract spatial correlations, where the number of groups is 30 and a sepconv block includes $1\times 1$  pointwise convolution and $3\times 3$ depthwise convolution. Note that there is no activation function in these two convolutions. After the 1×1 convolutional layer of sepconv block1, considering domain knowledge, we insert an Absolute Activation (ABS) layer to make our network learning the symbol symmetry of the residual noise. We also use residual connections in the two sepconv blocks, for accelerating network convergence, preventing gradient vanishing/exploding and improving classification performance.

\subsubsection{Experimental Verification and Analysis}
\par In order to compare Zhu-Net with Yedroudj-Net, we visualize the feature map in the first convolutional layer (the feature map of the CNN is difficult to interpret and visualize when the layer is deeper). The feature map is a good description of the feature extraction process. 
\par Both Yedroudj-Net and Zhu-Net are trained using WOW, at the payload of 0.2 bpp. We visualize the feature map of the first convolutional layer for Yedroudj-Net and the feature map of the sepconv block 2 for our network. The comparisons of the feature maps of stego and cover are shown in the Fig. 6.

\begin{figure*}
	\centering
	\includegraphics[width=0.5\textwidth]{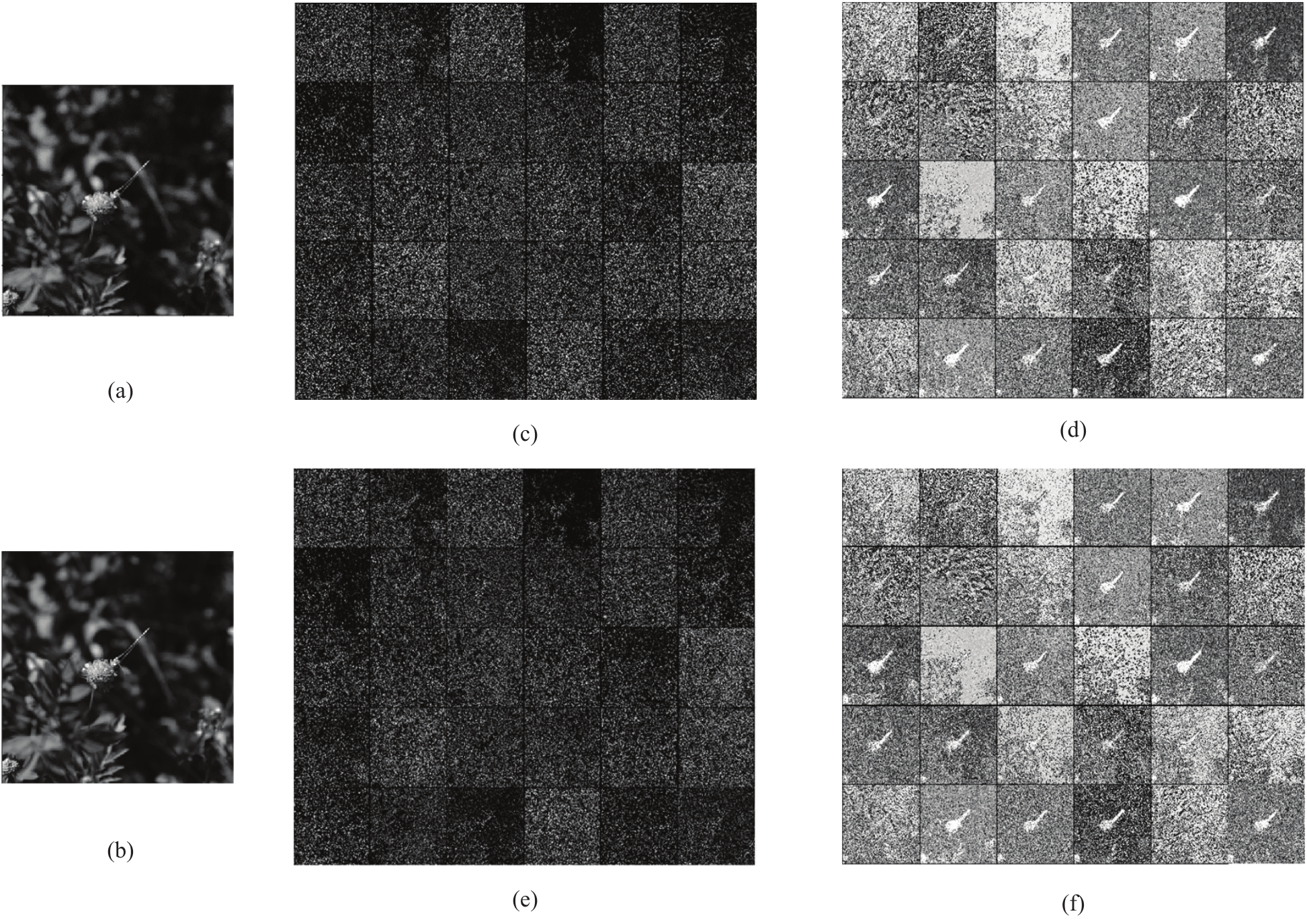}
	\caption{The comparison of feature maps between Zhu-Net and Yedroudj-Net. (a) Cover image. (b) Stego image. (c) The feature map of cover generated by Zhu-Net. (d) The feature map of cover generated by Yedroudj-Net. (e) The feature map of stego generated by Zhu-Net. (f) The feature map of stego generated by Yedroudj-Net.}

\end{figure*}

\par It is observed that the feature maps generated by the proposed scheme remain less image content information, and keep an increasing signal-to-noise ratio between stego signal and image signal. For either a cover or a stego, the proposed scheme can extract features with strong expression ability. Meanwhile, the similarity between every feature maps is relatively low, so that sepconv blocks facilitates subsequent convolution and classification. Comparative speaking, the feature maps generated by Yedroudj-Net remain more image contents, and the differences among feature maps are not obvious. 
\begin{table}
	\centering
	\caption{Steganalysis error rates comparison between Yedroudj-Net and Zhu-Net against two steganography algorithms WOW and S-UNIWARD at 0.2 bpp and 0.4 bpp. Both networks are trained and tested on BOSS dataset.}
	\begin{tabular}{ccc}

		\hline
		Algorithms        & Yedroudj-Net & Zhu-Net        \\
		\hline
		WOW(0.2bpp)       & 0.278        & \textbf{0.233} \\
		WOW(0.4bpp)       & 0.141        & \textbf{0.118} \\
		S-UNIWARD(0.2bpp) & 0.367        & \textbf{0.285} \\
		S-UNIWARD(0.4bpp) & 0.228        & \textbf{0.153} \\
		\hline
	\end{tabular}
\end{table}

\par In addition, we compare the detection error rate of Zhu-Net and Yedroudj-Net. Table III shows the performance of these two CNN networks against two steganography schemes such as S-UNIWARD and WOW. Experiment results show that, Zhu-Net obviously achieves better performance compared with Yedroudj-Net, reducing the detection error rate by 2.3\%-8.2\%.

\par Table IV shows the performance of detection error rate of Zhu-Net and Zhu-Net with full sepconv block against the algorithm WOW. The experiment results show that the detection accuracy goes down when all basic blocks are replaced by sepconv blocks. But the accuracy of Zhu-Net with all sepconv block is still better than Yedroudj-Net at a low embedding rate (e.g., 0.2 bpp). How to embed more sepcov blocks in CNN requires follow-up studies. Now, we choose the network with two sepconv block in our implementation so as to achieve a good detection performance.
\begin{table}
	\centering
	\caption{Steganalysis error rates comparison using Zhu-Net with different numbers of sepconv blocks against WOW at 0.2 bpp and 0.4 bpp. Both networks are trained and tested on BOSS dataset.}
	\begin{tabular}{ccc}

		\hline
		Algorithms& \tabincell{c}{Zhu-Net with \\ full sepconv blocks}& \tabincell{c}{Zhu-Net with \\two sepconv blocks}\\
		\hline
		WOW(0.2bpp) & 0.249 & \textbf{0.233} \\
		WOW(0.4bpp) & 0.152 & \textbf{0.118} \\
		\hline
	\end{tabular}
\end{table}

\subsection{Spatial pyramid pooling module}
\begin{figure}
	\centering
	\includegraphics[width=0.5\textwidth]{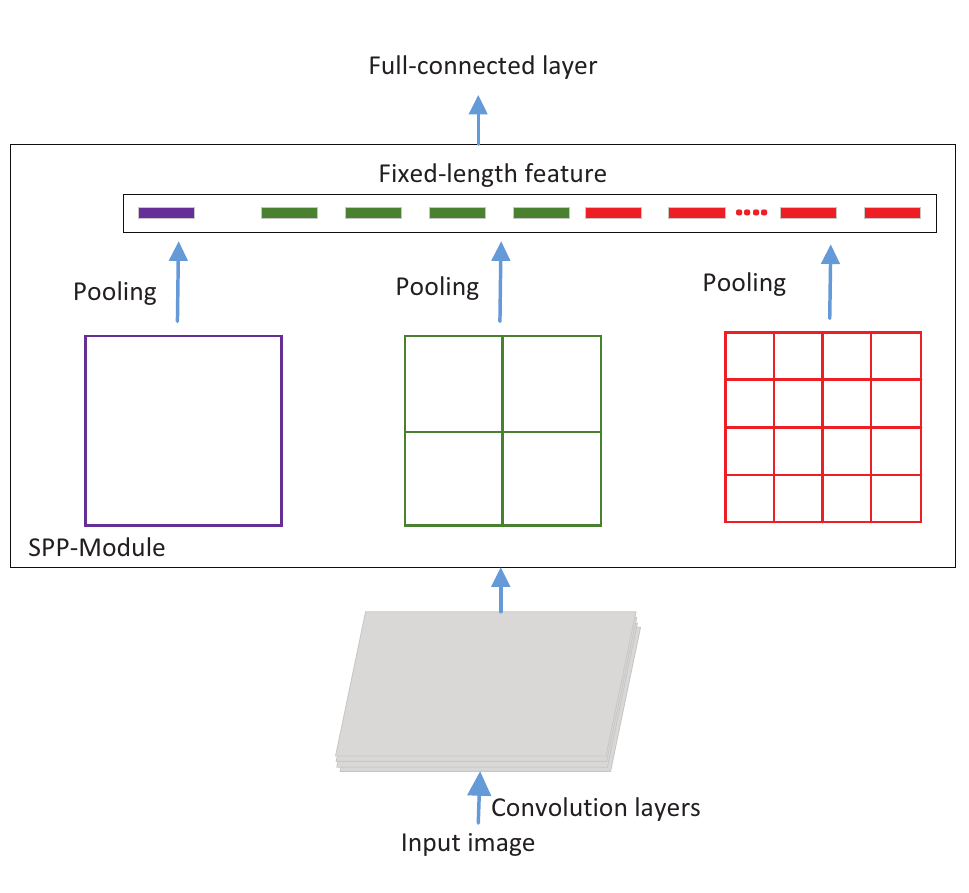}
	\caption{A network structure with a spatial pyramid pooling layer}
\end{figure}
\par For some steganalysis networks[18,21], a global average pooling (GAP) layer is added after the last convolution layer for down-sampling, which can greatly reduce the feature dimension. For image classification, GAP is generally used to replace full connected layer to prevent overfitting and reduce computational complexity. This global averaging operation equals to modeling the entire feature map, which leads to information loss of local features. However, for steganalysis networks, modeling local information is of key importance.
\par In our network, we use spatial pyramid pooling (SPP) to model local feature map, as shown in Fig. 7. SPP has the following properties[20]: 
\par (1) SPP outputs a fixed-length feature for any size input. 
\par (2) SPP uses multi-level pooling to effectively detect object deformation. 
\par (3) Since input is of arbitrary size, SPP can perform feature aggregation for any scale or size images.

\par Similar to [20], we divide the feature maps into several bins. In each spatial bin, we pool the responses of each feature map (we use average pooling hereinafter). The output of the spatial pyramid pool is a fixed $k\times M$ dimensional vector, where $M$ is the number of bins and $k$ is the number of filters in the final convolution layer. The major steps of the SPP-module mapping feature maps to fixed length vector are listed as follows:

\emph{\textbf{The steps of SPP-module mapping feature maps to fixed length vector}}
\par \textbf{Input:} The feature maps after basic block 4 with a size of $a\times a$ and channels of $K$. an $l$-level pyramid with $n\times n$ bins in each level.
\par \textbf{Output:} The fixed length feature with a size of $[1, K\times M]$, where $M$ is the number of bins.
\par \textbf{Step 1:} For a pyramid level of $n \times n$ bins, implement this pooling level as a sliding window pooling, where the window size win = $\left \lceil a/n \right \rceil$, and stride str = $\left \lfloor a/n \right \rfloor$ with  $\left \lceil \cdot \right \rceil$ and $\left \lfloor \cdot  \right \rfloor$ denoting ceiling and floor operations.
\par \textbf{Step 2:} Implement windows pooling on every feature map, obtain the generated feature with the length of $n\times n$.
\par \textbf{Step 3:} Repeat step1-step2 for every pyramid level in an $l$-lever pyramid.
\par \textbf{Step 4:} Stack all generated feature vectors together (in Pytorch we use torch.cat function). Pre-compute the length of each feature map by $M = \sum _{i=1}^{l} n \times n$, and the total length of feature is $K\times M$.

\par \textbf{Step 5:} Resize the output feature to a size of $[1, K\times M]$.

\par We use a 3-level pyramid pool ($4\times4$, $2\times2$, $1\times1$), which means that the number of bins is 21($4\times 4$ + $2\times 2$ + $1\times 1$). For a given size image, we pre-calculate the size of the output fixed-length vector. Assume that after the basic block 4 there is $a\times a$ (for example, $32\times 32$) size feature maps. When the pooling level is $4\times 4$, we divide the $32\times 32$ feature map into 16 small blocks, that is, the size of each small block is $8\times 8$. Then a GAP is performed on each $8\times 8$ block to obtain a 16-dimensional feature vector. In the pytorch toolbox, we can use average pooling (strinde:8, kernel:8) to achieve such sliding window pooling operations. The pooling level of $2\times 2$ and $1\times 1$ are similar. Finally, we can get a ($4\times 4 + 2\times 2 + 1\times 1) \times k$ dimensional vector, where $k$ is the number of filters in the last convolutional layer.
\par Interestingly, the $1\times 1$ level pooling actually equal to the global average pooling layer used in many steganalysis networks. This shows that we have gathered the information from feature maps at different levels, which not only integrates features of different scales but also better models the local features.

\par In order to verify the effectiveness of the SPP module in feature extraction, we compare Zhu-Net (with SPP-module) with Ye-Net and Yedroudj-Net (with GAP-module). All the networks are trained against WOW and S-UNIWARD, at the payload of 0.2 bpp. The experiment results are shown in Table V. Considering GPU computing power and time limitation, we construct a training set with two predefined sizes: $224\times 224$ and $256\times 256$. We resampled all the $512\times 512$ images to $256\times 256$ images and $224\times 224$ images. In order to compare with the existing networks, the size of testing images is still $256\times 256$.

\begin{table}
	\centering
	\caption{Steganalysis error probability comparison of Zhu-Net with different training schemes and Yedroudj-Net against the two algorithms WOW and S-UNIWARD at 0.2bpp. Both networks are trained and tested on BOSS dataset. }
	\begin{tabular}{ccc}

		\hline
		Algorithms& \tabincell{c}{WOW\\(0.2bpp)}& \tabincell{c}{S-UNIWARD\\(0.2bpp)}\\
		\hline
		\tabincell{c}{Ye-Net} & 0.331          & 0.400          \\

		\tabincell{c}{Yedroudj-Net}       & 0.278          & 0.248          \\
		\tabincell{c}{Zhu-Net wiht 256-size Tested}       & \textbf{0.234} & \textbf{0.281}          \\
		\tabincell{c}{Zhu-Net with multi-size Tested}        & 0.241          & 0.289 \\

		\hline
	\end{tabular}
\end{table}

\par The experiment results have shown that Zhu-Net has a higher accuracy than Yedroudj-Net and Ye-Net. That is, compared with the single-size training, the multi-size training can slightly improve the accuracy. We believe that the multi-size training relieves overfitting to a certain extent, and the multi-size dataset enhances the generalization ability of the network. 
\par Besides, we create a random-sized test set whose image sizes range from [224, 256]. The error rate of Zhu-Net against WOW and S-UNIWARD at 0.2 bpp is 0.241 and 0.289. 
\par The experiment results show that the detection with SPP-module is better than the detection with GAP, and the former network has better feature expression ability. Another advantage of using SPP-module is that it can handle inputs with arbitrary sizes.

\section{Experiments}

\subsection{The environments}
\par In our experiments, we use two well-known content-adaptive staganographic methods, i.e. S-UNIWARD [3], and WOW [2], by Matlab implementations with random embedding key. 
\par Our proposed CNN network is compared with four popular networks: Xu-Net[17], Ye-Net[19], Yedroudj-Net[21], and SRM + EC standing for the hand-crafted feature set named Spatial-Rich-Model [13] and Ensemble Classifier[32]. All the five networks are tested on the same datasets. All the experiments were run on an Nvidia GTX 1080Ti GPU card.

\subsection{Datasets}
\par In this paper, we use standard datasets to test the performance of the proposed networks. The two standard datasets are as follows:
\begin{itemize}
  \item the BOSSBase v1.01[33] consisting of 10,000 grey-level images of size $512\times 512$, never compressed, and coming from 7 different cameras.
  \item the BOWS2[34] consisting of 10,000 grey-level images of size $512\times 512$, never compressed, and whose distribution is close to BOSSBase.
\end{itemize}
\par Due to our GPU computing power and time limitation, we do all the experiments on images of $256\times 256$ pixels. The specific training set and test set division will be detailed in Section IV-D.
\subsection{Hyper-parameters}
\par We apply a mini-batch stochastic gradient descent (SGD) to train the CNN networks. The momentum and the weight decay of networks are set to 0.9 and 0.0005 respectively. Due to GPU memory limitation, the mini-batch size in the training is set to 16 (8 cover/stego pairs). All layers are initialized using Xavier method[35]. Based on the above settings, the networks are then trained to minimize the cross-entropy loss. During the training, we adjust the learning rate as follows (initialized to 0.005).
\par When the training iteration equals to one of the specified step values, the learning rate will be divided by 5. Concretely, the learning rate will be decreased at epochs 50, 150 and 250 respectively. During later training, using a smaller learning rate can effectively reduce training loss and improve accuracy. CNN training is up to 400 epochs. Actually, we often stop training before 400 epochs to prevent over-fitting. That is, when cross-entropy loss on training set keeps decreasing but detecting accuracy on validation set begins declining, we stop training. We choose the best trained model on validation set.

\subsection{Results}
\subsubsection{Results without data augmentation}

\begin{table}
	\centering
	\caption{Steganalysis error rates comparison using Yedroudj-Net, Xu-Net, Ye-Net, and SRM+EC against two steganography algorithms WOW and S-UNIWARD at 0.2 bpp and 0.4 bpp. All networks are trained and tested on BOSS dataset.}
	\begin{tabular}{ccccc}

		\hline
		Algorithms& \tabincell{c}{WOW\\(0.2bpp)}& \tabincell{c}{WOW\\(0.4bpp)}& \tabincell{c}{S-UNIWARD\\(0.2bpp)}& \tabincell{c}{S-UNIWARD\\(0.4bpp)}\\
		\hline
		SRM+EC       & 0.365          & 0.255          & 0.366          & 0.247          \\
		Xu-Net       & 0.324          & 0.207          & 0.391          & 0.272          \\
		Ye-Net       & 0.331          & 0.232          & 0.400          & 0.312          \\
		Yedroudj-Net & 0.278          & 0.141          & 0.367          & 0.228          \\
		Zhu-Net      & \textbf{0.233} & \textbf{0.118} & \textbf{0.285} & \textbf{0.153} \\
		\hline
	\end{tabular}
\end{table}

\begin{figure}
	\centering
	\includegraphics[width=0.5\textwidth]{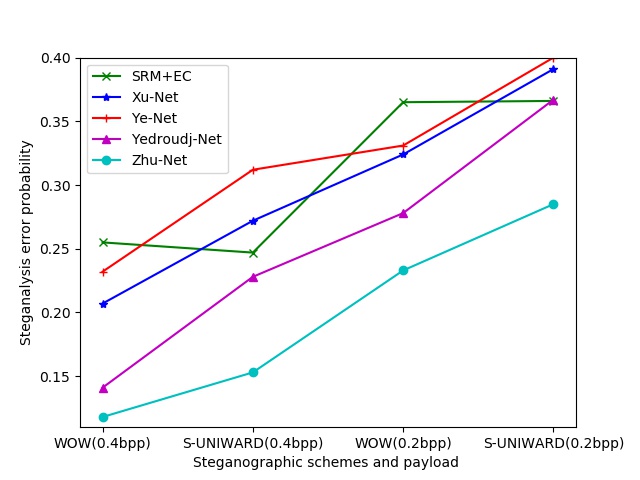}
	\caption{Steganalysis error rates comparison of the five steganalysis methods against two algorithms WOW and S-UNIWARD at 0.2 bpp and 0.4 bpp. All networks are trained and tested on BOSS dataset.}
\end{figure}
\par In Table VI, we report the performance comparison among steganalyzers without data augmentation. The BOSSBase images were randomly split into a training set with 4,000 cover and stego image pairs, a validation set with 1,000 image pairs, and a testing set containing 5,000 image pairs. For a fair comparison, we report the performance of Yedroudj-Net, Ye-Net, Xu-Net, and the Spatial Rich Model + the Ensemble Classifier (SRM + EC), against the embedding algorithm WOW and S-UNIWARD at payload 0.2 bpp and 0.4 bpp.
\par As Fig. 8 shows, the proposed network has significantly better performance than the other networks, regardless of the embedding method and payload. Due to the ability of CNN feature extraction, the proposed network has reduced error rate by 8.1\% to 13.7\%, comparing with the traditional network SRM+EC. Results also show that it is effective to use proposed network to optimize feature extraction and classification in a unified framework.
\par In addition, for S-UNIWARD and WOW with different payloads, the proposed network is 8.9\% to 11.9\% better than Xu-Net, 9.8\% to 15.9\% better than Ye-Net, and 2.3\% to 8.2\% better than Yedroudj-Net. It demonstrates that the proposed network effectively extracts the correlation of residuals, and has a good network structure including the multi-level pooling of SPP-module to improve the accuracy. Briefly, experiments prove that Zhu-Net outperforms other networks against various steganography schemes at any payloads. Note that above experiments were operated without tricks such as transfer-learning or virtual expansion of databases.

\subsubsection{Result with data augmentation}

\begin{table}
	\centering
	\caption{Steganalysis error rates comparison using Yedroudj-Net, Ye-Net and Zhu-Net on WOW at 0.2 bpp with a learning base augmented with BOWS2, and Data Augmentation}
	\begin{tabular}{cccc}

		\hline
		Algorithms & \tabincell{c}{BOSS} & \tabincell{c}{BOSS+BOWS2} & \tabincell{c}{BOSS+BOWS2+DA} \\
		\hline
		Ye-Net     & 0.331               & 0.261                     & 0.222                        \\
		Yedroudj-Net     & 0.278               & 0.237                     & 0.208                        \\

		Zhu-Net    & \textbf{0.233}      & \textbf{0.178}            & \textbf{0.131}               \\
		\hline
	\end{tabular}
\end{table}

\begin{table}
	\centering
	\caption{Steganalysis error rates comparison using Yedroudj-Net, Ye-Net and Zhu-Net on S-UNIWARD at 0.2 bpp with a learning base augmented with BOWS2, and Data Augmentation}
	\begin{tabular}{cccc}

		\hline
		Algorithms & \tabincell{c}{BOSS} & \tabincell{c}{BOSS+BOWS2} & \tabincell{c}{BOSS+BOWS2+DA} \\
		\hline
		Ye-Net     & 0.400               & -                         & 0.335                        \\
		Yedroudj-Net     & 0.366               & 0.344                     & 0.311                        \\

		Zhu-Net    & \textbf{0.285}      & \textbf{0.243}            & \textbf{0.171}               \\
		\hline
	\end{tabular}
\end{table}

\begin{table}
	\centering
	\caption{Steganalysis error rates comparison using Yedroudj-Net, Ye-Net and Zhu-Net on WOW at different payloads with Data Augmentation}
	\begin{tabular}{ccccc}

		\hline
		Algorithms& \tabincell{c}{Payload\\(bpp)}& \tabincell{c}{Ye-Net[13]}& \tabincell{c}{Yedroudj-Net[31]} & \tabincell{c}{Zhu-Net}\\
		\hline
		\multirow{4}*{WOW} & 0.1 & 0.348 & 0.330 & \textbf{0.233} \\
		                   & 0.2 & 0.262 & 0.208 & \textbf{0.131} \\
		                   & 0.3 & 0.225 & 0.189 & \textbf{0.084} \\
		                   & 0.4 & 0.184 & 0.158 & \textbf{0.065} \\

		\cline{1-5}

		\multirow{4}*{S-UNIWARD}
		                   & 0.1 & 0.400 & 0.383 & \textbf{0.268} \\
		                   & 0.2 & 0.335 & 0.331 & \textbf{0.171} \\
		                   & 0.3 & 0.256 & 0.221 & \textbf{0.125} \\
		                   & 0.4 & 0.226 & 0.171 & \textbf{0.081} \\

		\hline
	\end{tabular}
\end{table}

\begin{figure*}
	\centering
	\includegraphics[width=1\textwidth]{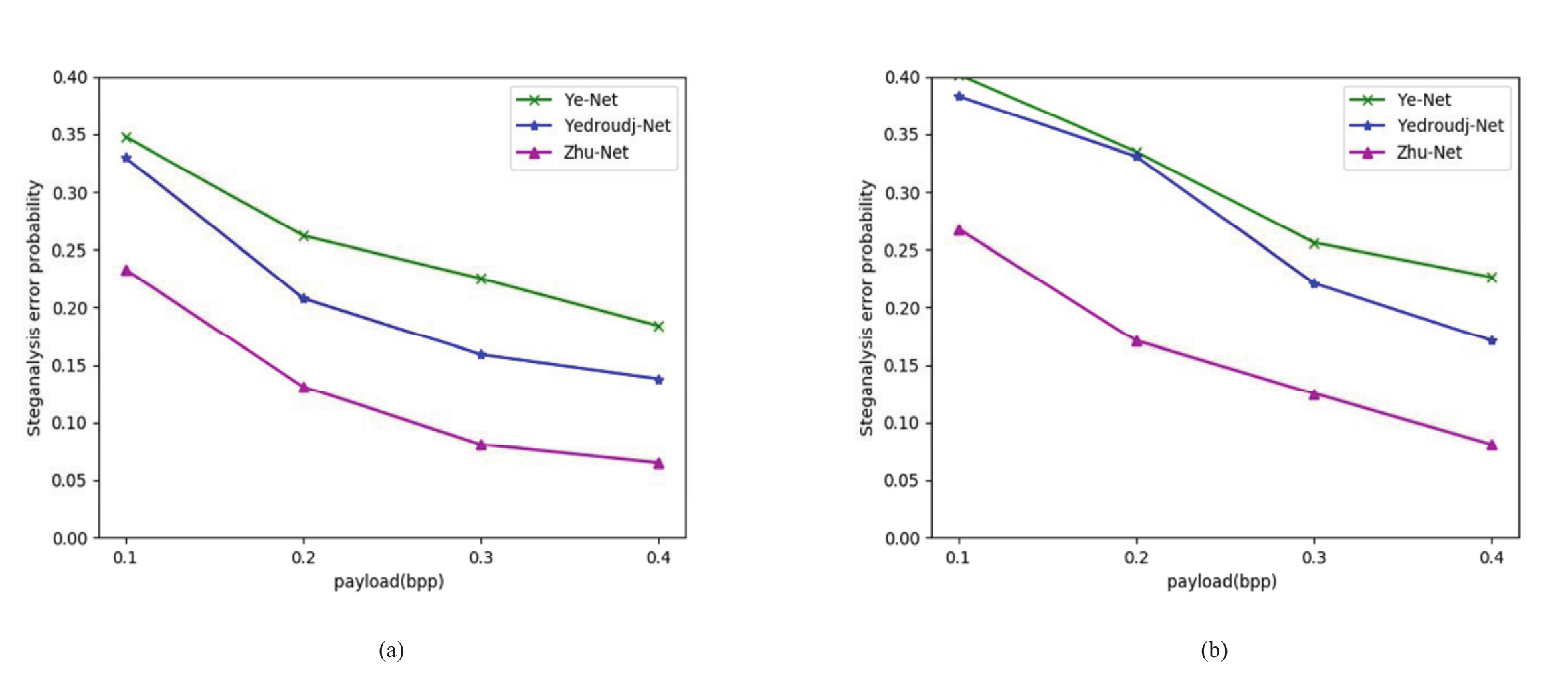}
	\caption{Steganalysis error rates comparison using YedroudjNet, Ye-Net and Zhu-Net on S-UNIWARD and WOW at different payloads. (a)WOW (b) S-UNIWARD}

\end{figure*}

\par Data augmentation can effectively improve the performance of the network by increasing the size of training database. Using a large database can improve accuracy and avoid overfitting. But, traditional data augmentation solutions such as clipping and resizing are not good choices for steganalysis because these solutions will destroy the correlation of pixels and drastically reduce the performance of network.
\par In order to study the effect of increasing datasets on the performance, we use the following data enhancement schemes. All photos are resampled into the size of $256\times 256$ pixels (using "imresize()" function in Matlab with default settings);
\par (1) training set BOSS: The BOSSBase images were randomly divided into a training set with 4,000 cover and stego image pairs, a validation set with 1,000 image pairs, and a testing set containing 5,000 image pairs.
\par (2) training set BOSS+BOWS2: Based on training set BOSS, 10,000 additional pairs of cover/stego pair (obtained by resampling BOWS2Base[36]) were added to the training set. The training database now contains 14,000 pairs of cover/stego images and the validation set contains 1,000 pairs from BOSS. 
\par (3) training set BOSS+BOWS2+DA: The database BOSS+BOWS2+DA is virtually augmented by performing the label-preserving flips and rotations on the BOSS+BOWS2 training set. The size of the BOSS+BOWS2 training set is thus increased by a factor of 8, which gives a final learning database made of 112,000 pairs of cover/stego images. The validation set contains 1,000 pairs from BOSS.
\par (4) testing set BOSS: it contains the remaining 5,000 images in BOSSbase other than the ones in training set BOSS. 

\par Table VII and Table VIII shows the comparisons of Yedroudj-Net, Ye-Net and Zhu-Net trained on different training sets, against the embedding algorithm WOW and S-UNIWARD at payload 0.2 bpp. The experiment results show that when the training set is incremented, the detection performance for all the network will be improved compared with that using the BOSS training set only. For WOW at 0.2 bpp, using training set BOSS+BOSW2 comparing to using only BOSS training set, Zhu-Net reduced the error rate by 5.5\% and achieved best results in all counterparts. The Yedroudj-Net and YeNet error rates are reduced by 4.1\% and 7\%. Similarly, for S-UNIWARD at 0.2 bpp, the detection error rates of the Ye-Net, Yedroudj-Net, Zhu-Net decreased by 2.2\% and 3.6\% comparing to only using BOSS training dataset, respectively. Zhu-Net still achieved best performance in all counterparts. The result shows that over-fitting is effectively mitigated by data augmentation.
\par This prompted us to use larger datasets for training. We further train three networks on BOSS + BOWS2 + DA. The results show that all CNN-based methods have improved performance. Compared those training only using BOSS, Zhu-Net still achieves best performance such as the decreased detection error by 10.2\% and 11.4\% against WOW and S-UNIWARD (Ye-Net by 10.9\% and 6.5\%, and Yedroudj-Net by 7\% and 5.5\%).
\par In Table IX and Fig. 9, we further illustrate the detection errors of three CNN-based steganalyzer against WOW and S-UNIWARD at different payloads. We note that Zhu-Net achieves significant improvements and best results compared with other CNN-based networks on different datasets against various steganography algorithms. Similarly, we attribute this improvement to the good network structure of Zhu-Net including sepconv block and SPP-module. 
\par All the experiments have shown that in order to effectively perform feature extraction and classification, CNN requires enough samples for training, even 112,000 pairs of pictures may not be enough. How to further increment dataset to meet the requirement of the steganalysis tasks needs further research.

\section{Conclusion}

\par It is significantly superior for steganalysis researchers to use CNN instead of traditional handcraft features - the ensemble classifier trained on the Rich Model. In this paper, we focus on designing a new CNN structure for steganalysis. The proposed network achieves a great improvement compared with existing CNN-based networks. The advantages of proposed network focus on: (1) We improve convolution kernel in preprocessing layer to extract the image residuals. Better convolution kernels reduce number of parameters and model local features; (2) We use separable convolution to extract channel correlation and spatial correlation of residuals, and thus the image content is removed from the features and the signal-to-noise ratio is improved. Utilization of residual in the preprocessing layer is more effective; (3) We use SPP-module instead of global pooling layer. By using different levels of average pooling to obtain multi-level features, the network performance is improved. Meanwhile, the SPP-module is a flexible solution for handling different sizes. It can map feature maps to a fixed number of dimensions, enabling detecting arbitrary sized images without any loss of accuracy. Finally, the performance of the proposed CNN is further boosted by using larger data sets. Experiment results show that the proposed CNN network is significantly better in the detection accuracy compared with the other networks.


\ifCLASSOPTIONcaptionsoff
  \newpage
\fi

\end{document}